\documentstyle[aps,multicol,prl]{revtex}
\begin{document}

\title{Attractions between charged colloidal spheres mediated by
correlated absorbed ions}
\author{Francisco J. Solis$^1$ and Monica Olvera de la Cruz$^2$ \\
$^1$ Physics Department, Hiram College, Hiram OH, 44238 \\
 $^2$ Department of Materials Science
and Engineering,  Northwestern University, Evanston IL 60208 \\
e-mail: $^1$ solisfj@hiram.edu, $ ^2$ m-olvera@northwestern.edu}
\maketitle

\maketitle
\begin{abstract}
We describe arrangements of ions capable of producing short-range
attractive interactions between pairs of charged colloidal
spheres in the low temperature strongly correlated limit. For
particles of radius $R$ with bare charge $Z$ and comparable
absorbed charge $-N$ ($N \sim Z$), the correlations contribution
to the spheres self-energy scales as $N^{3/2}/R$, and as $N/R$ for
the interaction energy between two touching spheres. We show that
the re-arrangement of charges due to polarization plays an
insignificant role in the nature and magnitude of the interaction.
\end{abstract}
\begin{multicols}{2}
%

 Electrostatic attractions between colloids have been reported in a large
number of circumstances \cite{larsen}. Experiments in colloidal
suspension confined into a small (almost two dimensional) region
\cite{fraden,grier1,grier2} have shown direct evidence of these
attractions. Simulations of colloidal suspensions  in bulk
\cite{kremer2,prl1} and confined systems \cite{all2} have also
reported electrostatic attractions. The simulations have
explored, in particular, the strongly correlated limit in which
the thermal energy of the system is smaller than the typical
interaction energy between the charges absorbed to the colloids.

Recent studies on colloidal suspensions have elucidated the effect
of depletion forces \cite{all2} and of Van der Waals interactions
\cite{hansen}, as well as the importance of including
correlations \cite{kremer2,prl1,pincus,levin,gonzalez} to predict
attractions. In this letter we derive analytic expressions for the
interaction energy between two colloids in the strongly
correlated limit by considering the distribution of absorbed
counterions around them  as they approach. Though correlations
between absorbed counterions are known to induced short range
attractions between charged surfaces \cite{marcelja,rouzina}, and
between aligned rods \cite{rods,shkrods,arenzon}, it has been
noted \cite{hansen} that the geometry change to colloids is not
trivial.

Systems involving large charged objects (polyions) whose charge
is compensated by mobile low-valence ions can be study by
considering first the electrostatic ground state of a subsystem
that consists of one or few polyions with a fixed number of
counterions. Once the ground state is constructed, is possible to
draw conclusions about the system by introducing thermal
fluctuations and setting the subsystem in contact with the rest of
the system which then acts as a thermal and particle reservoir.
This approach is justified by the fact that there is a small
region surrounding the polyion in which the electrostatic
interactions are so dominant that thermal fluctuations are only a
minor correction. It has been used, for example, to study the
interactions between charged surfaces \cite{rouzina} and between
charged linear polyelectrolytes \cite{rods,shkrods}, the
conditions of solubility of DNA and other linear polyelectrolytes
in multivalent salt solutions \cite{collapse,shksalt}, and to
describe the differences between charging colloids or surfaces
with monovalent or multivalent counterions \cite{shkover}. Results
from these theories were used in Ref. \cite{kremer2} to obtain
simple but accurate predictions that explain many colloidal
simulations results.

While in principle it might be possible to solve the ground state
of a system consisting of one charged sphere and a set of $N$
absorbed point-charges, such a ground state has to be solved
almost independently for each different $N$. It would be even
more difficult to obtain exact results for the case of two
spheres, although numerical results can be obtained for $N$ not
too large. Instead of attempting such solutions, we propose an
approximate solution that satisfies a number of desirable physical
conditions. We first decompose the charge distribution on the
sphere into a multipolar series for each sphere. We then truncate
the series and propose values for the dominant terms. Using this
ansatz, we obtain an approximation for the interaction energy
between two spheres.

Our main results are simple to understand. We recover the expected
negative self-energy for a single charged sphere due to the
discrete nature of the absorbed counterions, which is of the form
$-N^{3/2}/R$ \cite{kremer2}. Indeed, since $R \sim N^{1/2} a$,
where $a$ is the typical distance between the absorbed charges,
this self-energy is roughly proportional to $-N/a$, as in the
case of a flat surface. If the spheres have bare charge $Z$, and
$N$ absorbed (opposite) charges, the effective charge is $Q=Z-N$,
and thus there is an  interaction energy between two spheres at a
distance $r$ of the form $Q^2/r$, but there are important
contributions arising from the particle nature of the absorbed
charge. When the spheres are in contact there is a region in
which the charges from both spheres are strongly correlated. In
this region the bare and absorbed charge density are doubled and
the typical distance between charges is now of the order of
$a/\sqrt{2}$. The charges in both spheres remain correlated up to
when the separation between corresponding points in the spheres
are separated by a distance of the order $a$. This defines an
effective area of contact between the spheres given by $\pi R a$,
that contains $Na/4R$ charges. The interaction energy is then
{\it negative} and of order $N/R$. Thus, there can be a very large
attractive component to the interaction between the spheres
capable of producing a net attraction between them for almost
neutral spheres ($N \sim Z$).

%
%

We assume that the total bare charge of the colloidal sphere $Z$
is uniformly distributed on its surface or, equivalently,
concentrated at the center of the sphere. The surface charge
density of the N absorbed particles is then written as $\rho({\bf
r})=\Sigma_{i}{\delta( {\bf r}- {\bf r}_{i})} $ where ${\bf
r}_{i}$ are vectors pointing from the center of sphere to a
position at the surface. The multipolar moments $M_{n}^{m}$ for
this distribution are
\begin{equation}
M_{n}^{m}=\sqrt{\frac{4\pi}{2n+1}}\int d{\bf r}
Y_{n}^{m}(\hat{{\bf r}})r^{n} \rho({\bf r}) \label{moments}
\end{equation}
 where
$Y_{n}^{m}$ are spherical harmonics. The charge density can be
then re-expressed as the multipolar expansion:
\begin{equation}
\rho({\bf r})= \sum_{n,m} \sqrt{\frac{2n+1}{4\pi}}
M_{n}^{m}\frac{\delta(r-R)}{R^{n+2}}Y_{n}^{m}(\hat{{\bf r}}).
\label{surface}
\end{equation}
Using Maxwell's representation \cite{maxwell} of the spherical
harmonics it is also possible to replace the surface density by a
singular distribution placed at the center of the sphere given by
\begin{equation}
\rho({\bf r})= \sum_{n,m} \frac{(-1)^{n-|m|}M_{n}^{\pm
m}\partial_{z}^{n-|m|}(\partial_{x}\mp
i\partial_{y})^{|m|}\delta({\bf r})}{((n-m)!(n+m)!)^{1/2}}.
\label{delta}
\end{equation}
We will use the more compact notation $\rho=\sum
M_{n}^{m}\{n,m\}$ where the brace $\{n,m\}$ stands in for the
suitable functional form in this expression. The potential
generated by this charge density outside of the sphere, at $r>R$,
is
\begin{equation}
\phi({\bf r})=\Sigma_{n,m}\sqrt{\frac{4
\pi}{2n+1}}M_{n}^{m}Y_{n}^{m}(\hat{{\bf r}})\frac{1}{r^{n+1}}.
\end{equation}

In the low temperature limit we expect the charges to occupy
positions that up to permutations and rotations of the whole
sphere are uniquely determined. For large $N$, the
inhomogeneities arising from the particle nature of the charges
in this static ground-state distribution should create only
negligible contributions for the first few multipolar moments
other than $n=0$, with the next non-vanishing moments arising at
$n$ values that start to resolve distances of order $a$, namely
$n \sim N^{1/2}$. To make analytical calculations feasible we
truncate the multipolar expansion and represent the sum of the $N$
delta functions as a smooth function with $N$ clearly
distinguished maxima. Since the interaction between the absorbed
charges is repulsive one can expect four nearest neighbors per
particle in the ground state and we also require this from our
truncated expansion. Our ansatz is then a truncation of the
series Eq.(\ref{surface}) that only consider moments with $n$
values up to $n=3l/2$ with $l=\sqrt{2N}$. Further, within this
range we take only four moments to be non zero:
\begin{eqnarray}
 \rho=-N\{0,0\}&+&\frac{(2N)^{3/4}R^{l}}{\pi^{-1/2}}\{l,0\}+ \nonumber \\
&&\frac{2^{7/8}N^{5/8}R^{l}}{\pi^{3/4}}(\{l,l\}+\{l,-l\}).
\label{proposed}
\end{eqnarray}
The precise values of the moments given here are discussed below.

The proposed charge density Eq.(\ref{proposed}) has $N$ maxima,
distributed between $l=\sqrt{N/2}$ meridians generated by the
$m=l$ terms.  Each meridian has $l/2$, non uniformly
distributed maxima generated by the $m=0$ term. 
A consistent extension of our ansatz would give the next non-zero
moments at $n$ and $\pm m$ values that are integer multiples of
$l$. In the limit of infinite radius and near the equator of the
sphere the distribution becomes equivalent with the description,
by a few Fourier modes, of a square lattice of charges on a
plane. This limit is indeed the inspiration for our choice. The
major inaccuracies of this distribution are near the poles where
the maxima are closer than they should be.

For large $N$, and near the equator, the spherical harmonic
$Y_{l}^{0}$ has an oscillatory form
${\pi}^{-1}\cos(l\theta+\psi)$, with $\theta$ the polar coordinate
and $\psi$ a phase. If we make the position of the charges to
coincide with the peaks of this approximate distribution, we
obtain, using Eq.(\ref{moments}),
$M_{l}^{0}=\pi^{-1/2}(2N)^{3/4}R^{l}$. Again, near the equator,
we have that $Y_{l}^{\pm l} =e^{\pm il\phi} f(\theta)$, where
$\phi$ is the azimuthal angle, and $f$ is a multiple of the
Legendre function $P^{l}_{l}$ that has a maximum in the equator
and is important only in a narrow region. We assume that the
particles positions coincide with the maxima of the real parts of
the waves in the $\phi$ coordinate while the sum over their
contributions along the $\theta$ is approximated by a multiple of
the integral $ \int f(\theta)d\theta$ from which $M_{l}^{l}=
2^{7/8}(N)^{5/8}R^{l}/(\pi^{3/4})$.
%
%

To discuss the interaction between spheres we must consider the
redistribution of charges that occurs when they are put in close
contact with each other. To first order we can represent this
redistribution as a flow generated by the projection, onto the
sphere, of a uniform displacement of charges in the direction of
the polarization. If the displacement of the charges is ${\bf
d}$, the new charge distribution is approximately $\rho({\bf r})
+\Delta \rho({\bf r})=\rho({\bf r})+{\bf d}\cdot \nabla \rho({\bf
r})$. Using the representation Eq. (\ref{delta}) and assuming the
displacement to be in the ${\hat {\bf x}}$ direction, it can be
shown that the displacement applied to our ansatz
Eq.(\ref{proposed}) gives rise to new non-zero extra moments:
\begin{eqnarray}
\Delta
\rho&=&\frac{P}{\sqrt{2}}\{1,1\}+\frac{PlM_{l}^{0}}{2N}\{l+1,1\}+
\frac{PlM_{l}^{l}}{N}
\{l+1,l+1\}\nonumber \\
&&+ \frac{PM_{l}^{l}}{\sqrt{2}N}\{l+1,l-1\}+ (m\rightarrow
-m).\label{polar}
\end{eqnarray}
This equation displays only the positive $m$ terms. The magnitude
of the generated dipole moment is $P=dN$, but as this expression
shows, the flow of charges also affects the higher order
multipoles. The linear approximation is valid only if the average
displacements of the charges are smaller than the typical distance
between them $d<a$

%
%
The energy of interaction between the absorbed charges and the
colloid is $-ZN/R$. If the decomposition of the ground state
configuration of the absorbed charges into multipolar moments
$M_{n}^{m}$ is known, their self energy is $E=\sum_{n,m}
M_{n}^{m}M_{n}^{-m}/{2R^{2n+1}}$. This expression is infinite
since it contains the self-energy of every absorbed charge. The
harmonic decomposition of a single point charge shows that its
self-energy splits into equal contributions of magnitude $1/2R$
for each $n$ number giving, symbolically,
$E_{single}=(1/2R)\sum_{n} 1$. Subtracting  $N$ times this amount
from the energy at each $n$ level we obtain:
\begin{equation}
E=-\frac{ZN}{R}+\sum_{n} \left(
-\frac{N}{2R}+\sum_{m}\frac{M_{n}^{m}M_{n}^{-m}}{2R^{2n+1}}
\right).
\end{equation}
Evaluating the energy of our ansatz for the charge distribution
and including the possibility of a charge redistribution with
dipole moment $P$ the main terms in the evaluation of the energy
are
\begin{equation}
E= \frac{Z^2-2ZN+N(N-1)}{2R} -\alpha \frac{N^{3/2}}{R}+\beta
\frac{P^2N^{1/2}}{2R^3}, \label{self}
\end{equation}
where $\alpha=(3/2^{3/2})-1/(\pi 2^{1/2})$, and $\beta=1/(2^{1/2}
\pi)$. In the second and third terms we have retained only the
contributions with highest powers in $N$.  The monopole
contributions have been explicitly separated. In the absence of
polarization, this result is clearly consistent with the expected
behavior of the self-energy in the $N \rightarrow \infty$ limit
that is equivalent to charges in a flat surface in which the
self-energy scales as $N(1/a) \sim N (N/R^2)^{1/2}$.

To consider two spheres we orient the  reference systems for each
of them  so that their centers lie in the x-axis of both systems.
The total distance between centers is $r$. The interaction energy
can be written as the integral of the potential generated by one
sphere $\phi$ times the charge density of the second sphere
$\rho'$, $E=\int d{\bf r} \phi \rho'$. Writing down both $\phi$
and $\rho'$ in their multipolar expansions and using the Maxwell
representation, the integral becomes a sum of terms that are
products of derivatives of $1/r$ times derivatives of a delta
function at the center of the second sphere. Repeated integration
by parts leads to
\begin{equation}
E=\sum_{n,m,p,q}E_{n,m,p,q}\frac{M_n^m M'^q_p}{r^{n+p+1}}
\end{equation}
where the coefficient is, using $s=n+m,t=n-m,s'=p+q,t'=p-q,
S=s+s',T=t+t'$,
\begin{equation}
E_{n,m,p,q}=\left[\frac{S!T!}
{s!s'!t!t'!}\right]^{1/2}\sqrt{\frac{4
\pi}{2S+1}}Y_{(S+T)/2}^{(S-T)/2}(\frac{\pi}{2},0).
\end{equation}
The point of evaluation of the spherical harmonic reflects the
direction of the relative position of the spheres in the chosen
coordinates, namely the $\hat{x}$ axis. It can be shown that for
large moment numbers the magnitude of the coefficient has its
maximum values when $n=p$ and $m=\pm q$, and decays away from
this situation as $\exp(-(s-(S/2))^2-(t-(T/2))^2)$. Therefore,
the main contributions to the energy from high order multipoles
appears when the multipoles in both spheres correspond, that is,
when the charge distributions are similar and interlock. In our
truncation we are able to pick the relative phases of the
multipoles in the spheres to get an attractive energy.

We will consider only the case of two spheres of equal radius.
Since we are interested in exploring the possibility of attractive
interactions between spheres we consider only the case in which
they touch $r=2R$. If attractions are present this will be their
optimal position. Considering the case of spheres that have equal
total charge and magnitude of polarization we obtain the following
interaction energy:
\begin{eqnarray}
E_{int}&=&\frac{(Z-N)^2}{2R}-\gamma
\frac{N}{2R}-\frac{(P-P')(Z-N)}{(2R)^2}+
 \nonumber \\
&& \frac{PP'}{(2R)^3}(-1+2\gamma) \label{interaction}
\end{eqnarray}
where $\gamma=\pi^{-2}+2^{3/4}\pi^{-5/2}$. The total energy is
obtained from adding the self-energy and the interaction energy.
Minimizing the total energy with respect to the polarization we
arrive to the conclusion that the polarization of the spheres
should be antiparallel ${\bf P}=-{\bf P}'$, and with magnitude
$P=RQ/\beta N^{1/2}$. The polarization is negligible in the limit
of large $N$. The reason for this is that a flow of mobile charges
towards the point of contact requires a large increase of energy
in a large depleted area, while the decrease in energy occurs
only in the small region of contact. Thus, we find that the ground
state is very rigid, and the interaction between spheres does not
alter much its local structure. The polarization corrections can
then be ignored and the interaction energy is simply the result
given in Eq.(\ref{interaction}) without polarization terms. If the
effective charge is small, the interaction is dominated by the
high multipole correlations and it scales as $N/R$. The rigidity
of the local structure makes this result robust when applied to
considerations of more than two spheres.

Given our result for the evaluation of the interaction energy, we
are ready to comment on the problem of attraction in more
realistic conditions. The Bjerrum number $B=e^2/a4\pi
\epsilon_{0}k_{B}T$ compares the electrostatic interaction between
charges at a distance $a$ with the typical thermal energy
$k_{B}T$. We can consider $z$-valent counterions and then, if we
have $z^2B>1$, the correlation energy can be very strong.
The exact number of charges in the spheres is controlled by the
equilibrium of the energy between absorbed charges and a chemical
potential $\mu$ for the extraction of them from the volume not
occupied by the colloidal spheres. To solve for the number of
absorbed charges we can minimize the free energy $F=E(N)-\mu N$
with respect to $N$. Short range attractions between two charged
colloids proportional to $N/R$ result from the "interlocking"
correlations of the thin layer of counterions at the surface of
the colloids in the area of contact. These correlations easily
arise, for high $Bz^2$, when the number of  absorbed counterions
is large and comparable to the bare charge. The relative location
of spheres influences the amounts of absorbed charges but their
specific values are mostly dependent on the interaction parameter
$Bz^2$ and on the chemical potential $\mu$. This suggests that it
is more likely to observe attractions in solutions of high valence
counterions or in systems where the accessible volume for the
counterions is small as in the case of colloids in confined
geometries. Though in principle the confinement of colloidal
suspensions by weakly charged surfaces imposes non-trivial
boundary conditions that affect the local electrostatic fields,
we can always analyze the local configurations of ions under
these externally imposed conditions. If the external conditions
are successful in creating attractions, the local arrangement of
charges are as we have described here.

The explanation for the presence of long-range attractions put
forward in Ref. \cite{kremer2} is that there are attractive
metastable states characterized by pairs of spheres with opposite
total charges. The overcharging of some spheres is possible due
to the negative correlation self-energy in each sphere
Eq.(\ref{self}). What our calculation shows is that even if the
effective charges are equal, the ground state is attractive. The
metastable states, on the other hand, reduce the potential
barrier to achieve close contacts,  and at the free energy level
might indeed provide a long range attractive interaction. At
short distances we expect the charge to redistribute to the type
of configurations discussed here.

We thank Cherry Murray for useful conversations. This work was
sponsored by the National Science Foundation, grants DMR9807601
and DMR9632472.

\end{multicols}

\begin{thebibliography}{5}
\bibitem{larsen}A. E. Larsen and D. G. Grier, { Nature} (London) {\bf 385},
230 (1997).
%
\bibitem{fraden}G. M. Kepler and S. Fraden,
{  Phys. Rev. Lett.} {\bf 73}, 356 (1994).
%
\bibitem{grier1}J. C. Crocker and D. G. Grier,
{  Phys. Rev. Lett.} {\bf 77}, 1897 (1996).
%
\bibitem{grier2}D. G. Grier, {  Nature} (London) {\bf 393}, 621
(1998).
%
\bibitem{kremer2} R. Messina, C. Holm, and K. Kremer,
  {  Phys. Rev. Lett. } {\bf 85}, 872 (2000).
(2000).
%
\bibitem{prl1} P. Linse and V. Lobaskin, {  Phys. Rev. Lett.}
{\bf 83} 4208 (1999).
%
\bibitem{all2}E. Allahyarov, I. D'Amico, and H. L\"{o}wen, {  Phys. Rev. E}
{\bf 60}, 3199 (1999).
%

\bibitem{hansen} R. van Roiji and J.-P. Hansen, {  Phys. Rev. Lett.}
{\bf 79} 3082 (1997).
%
\bibitem{pincus} N. Gronbech-Jensen, K. M. Beardmore and P. Pincus,
{Physica A} {\bf 261} 74 (1998).
%
\bibitem{levin} Y. Levin,
{Physica A} {\bf 265} 432 (1999).
%
\bibitem{gonzalez} P. Gonz\'alez-Mozuelos and  M.D. Carbajal-Tinoco, J.
 Chem. Phys.  {\bf 109} ,11074 (1998).
%
\bibitem{marcelja} R. Kjellander and S. Marcelja, {  Chem. Phys. Lett.}
{\bf 112} 49 (1984).
%
\bibitem{rouzina}I. Rouzina and V. Bloomfield,
{  J. Phys. Chem.} {\bf 100}, 9977 (1996).
%
\bibitem{rods} F. J. Solis  and M. Olvera de la Cruz,
{  Phys. Rev. E} {\bf 60}, 4496 (1999).
%
\bibitem{shkrods}B. I. Shklovskii, {  Phys. Rev. Lett.} {\bf 82}, 3268
(1999).


\bibitem{arenzon}
J. J. Arenzon, J. F. Stilck, and Y. Levin {European Physical
Journal} {\bf B12}, 79 (1999).
%
\bibitem{collapse}F. J. Solis and M. Olvera de la Cruz,
{  J. Chem. Phys.} {\bf 112}, 2030 (2000).

\bibitem{shksalt}  T. T. Nguyen, I. Rouzina, B. I. Shklovskii:
   J. Chem. Phys. {\bf 112}, 2562 (2000).

\bibitem{shkover} B.I. Shklovskii, {  Phys. Rev. E} {\bf 60}, 5802 (1999).

\bibitem{maxwell} See for example: Bateman Manuscript Project {\it Higher transcendental functions
vol.III},  (McGraw-Hill, New York, 1953)  Ch. 11.

%
%
\end{thebibliography}
\end{document}